# Surface structure determines dynamic wetting

Jiayu Wang[1], Minh Do-Quang[2], James J. Cannon[1], Feng Yue[1], Yuji Suzuki[1], Gustav Amberg[2], Junichiro Shiomi[1,3,*]

[1]Department of Mechanical Engineering, The University of Tokyo, Tokyo, Japan
[2]Department of Mechanics, Linné Flow Centre, The Royal Institute of Technology, Stockholm, Sweden
[3]CREST, Japan Science and Technology Agency, Tokyo, Japan

Liquid wetting of a surface is omnipresent in nature and the advance of micro-fabrication and assembly techniques in recent years offers increasing ability to control this phenomenon. Here, we identify how surface roughness influences the initial dynamic spreading of a partially wetting droplet by studying the spreading on a solid substrate patterned with microstructures just a few micrometers in size. We reveal that the roughness influence can be quantified in terms of a line friction coefficient for the energy dissipation rate at the contact line, and that this can be described in a simple formula in terms of the geometrical parameters of the roughness and the line-friction coefficient of the planar surface. We further identify a criterion to predict if the spreading will be controlled by this surface roughness or by liquid inertia. Our results point to the possibility of selectively controlling the wetting behavior by engineering the surface structure.

[*]shiomi@photon.t.u-tokyo.ac.jp

The rate of dynamic spreading of a liquid on a rough surface, whether during adhesion of a Gecko's feet to a surface[1], the self-cleaning of a Lotus leaf during rainfall[2] or the splashing of an object falling into liquid[3], depends on the degree of surface roughness. Understanding of wetting on a rough surface is important because even a *macroscopically flat* surface has microscale roughness. Thus there is recent growing interest in studying dynamic wetting on rough surfaces, supporting the development of emerging technologies such as ink-jet printing of electronics[4-7], boiling enhancement[8], droplet repulsion[9], material patterning and design[10-14], and adhesion[15].

The key issue is to identify the primary physical effect that resists the wetting. Tanner's law[16] assumes that viscous dissipation is the dominant source of resistance and uses a capillary time scale to derive a power-law time dependence between the spreading radius $r$ and time $t$ of $r \sim (t/\tau)^{1/10}$, where the capillary time is given by $\tau = \mu R/\gamma$ with $\mu$, $R$ and $\gamma$ the viscosity, droplet radius and surface tension, respectively. Recently however it was found that wetting at the initial stage is much faster than Tanner's law predicts[17-20]. For a completely wetting case ($\theta \sim 0°$) the spreading radius was shown to follow $r \sim (t/\tau)^{1/2}$, which can be derived by using the inertial time scale $(\rho R^3/\gamma)^{1/2}$ in analogy to droplet coalescence[17], where $\rho$ is the liquid density. Interestingly, the inertial scaling properly describes the dependence of spreading rate on the droplet size/wettability of the surface[18] although the exponent of the power law $r \sim (t/\tau)^\alpha$ is no longer 1/2 but decreases with increasing equilibrium contact angle.

Another branch of analysis discusses the resistance in terms of non-hydrodynamic energy dissipation at the contact line[21-25]. In Molecular Kinetic Theory (MKT)[22,23], wetting is described in terms of transition of molecules from one adsorption site to another with kinetic energy overcoming an activation barrier. The local non-hydrodynamic energy dissipation due to events on the molecular scale can be effectively described in terms of the contact line friction parameter[21] $\mu_f$, which has the same units as dynamic viscosity. This phenomenological parameter is defined such that the energy dissipation rate associated with molecular processes at the contact line, per unit length of contact line, is

$$w = \mu_f U^2 \tag{1}$$

One way to quantify the contact-line energy dissipation rate $\mu_f$ is to implement it into the boundary condition of Cahn-Hilliard Navier-Stokes (CHNS) simulations and find the value that makes the simulated spreading match that of corresponding experiments[24,25].

Despite the progress in understanding the dynamic wetting on flat surfaces, investigations of truly dynamic spreading on rough surfaces are infrequent. Cazabat and Stuart[26] undertook experiments on spreading of perfectly-wetting silicone oils on glass surfaces that had been roughened to various degrees and found that the presence of roughness could enhance spreading.

In a recent study, Yuan and Zhao[27] undertook experiments with silicone oil spreading on silicon substrates with pillars between 10 and 20 micrometers in size, and made illustrative molecular dynamics simulations, showing that for the completely wetting case the liquid may be pulled ahead through the forest of pillars, in front of the visible contact line. Kusumaatmaja et. al.[28] meanwhile studied the influence of anisotropy on the spreading. These works either concern late stage of complete wetting or feature comparatively large (>10 μm) posts. In this study however, we aim to identify the role of roughness in determining the dynamic wetting by investigating the initial stage of partial wetting on a surface with periodic microstructures of just a few μm in size. We show that the influence of the microstructure can be quantified by $\mu_f$, extracted by combined experimental and CHNS simulation analysis. Furthermore, by utilizing systematic variation of the pattern geometries and liquid properties, the underlying mechanism of the influence is identified.

Droplet wetting experiments were performed on flat and microstructured surfaces. The microstructures are periodic arrays of pillars with square cross sections of side length *a*, spacing *b*, and height *h*, (Fig. 1). The height of the pillars is kept constant at 1.6 μm while *a* is varied from 0.5 to 5 μm, and *b* from 1 to 45 μm. This is smaller than previous experiments[27,29,30] in an effort to achieve small-scale surface roughness. The surfaces of the flat and microstructured substrates are uniformly covered with native oxide layers. We have chosen a partially wetting liquid to realize Wenzel wetting throughout the study. The initial droplet radius is fixed to 0.5 mm. Static contact angle measurements find significant hysteresis (difference between advancing and receding contact angles)[31] that grows and eventually saturates with the normalized surface area or roughness parameter $S=1+4ah/(a+b)^2$ of the microstructure. Here the roughness is defined as the total area $(a+b)^2+4ah$ of the microstructured surface divided by the projected area $(a+b)^2$.

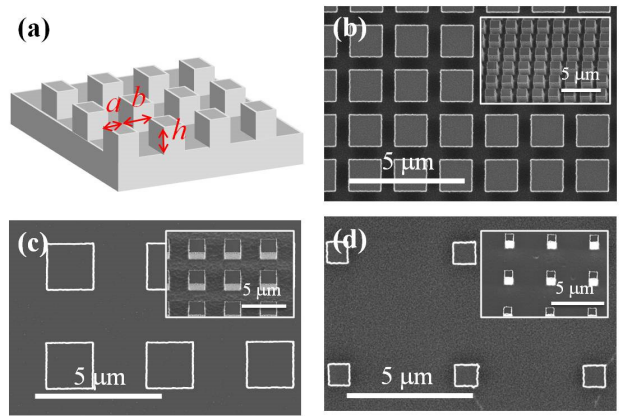

Fig. 1 (a) Schematic showing terminology for pillar width *a*, separation *b* and height *h*. (b-d) Scanning electron microscopy images of representative microstructured surfaces for (b) *b*/*a*=1/3, (c) *b*/*a* =1, and (d) *b*/*a* =4.

The snapshots of the droplet spreading [Fig. 2(a)] and the corresponding time histories of the spreading radius [Fig. 2(b)], taken at the initial stage of wetting before the upper half sphere deforms, clearly show that spreading is hindered by the microstructures. Furthermore, as shown in Fig. 2(c), the normalized average spreading rate decreases with $S$, indicating that the resistance is highly sensitive to the microstructure geometry. In case of the microstructure with the largest roughness $S$, the wetting speed is reduced to about half of that for the flat surface with the same surface chemistry. The figure also shows that the hindrance is larger for larger viscosity. Note that the wetting is expected to be axisymmetric in this early stage of wetting unlike the dynamic wetting at a later stage close to equilibrium, where the contact line shape is known to become non-circular following the asymmetry of the microstructures[32,33]. Absence of wicking was confirmed by observing the droplet from the top.

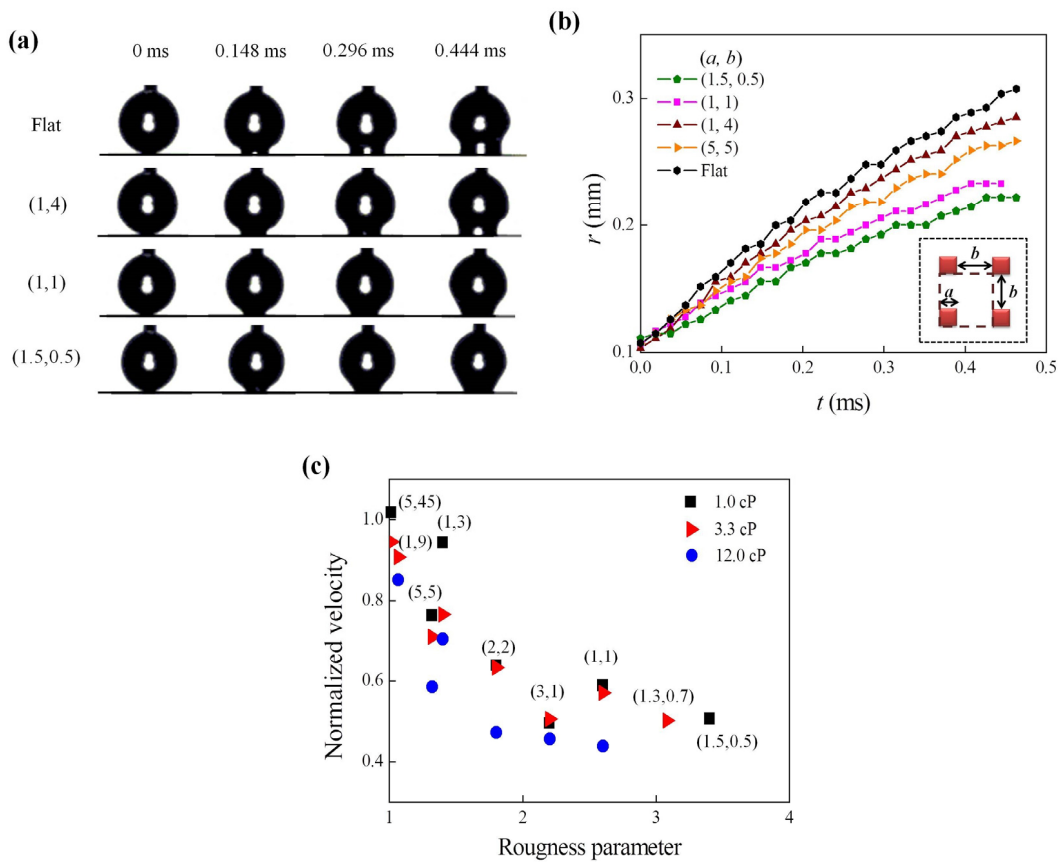

Figure 2 (a) Snapshots of spreading on representative substrates and (b) the evolution of the spreading radius for the 1.0 cP viscosity case. (c) Dependency of the average spreading rate on the roughness parameter $S$ and viscosity (1.0, 3.3, and 12.0 cP), normalized against that for the respective flat surfaces.

We will now quantify the influence of the microstructures in terms of the line friction coefficient $\mu_f$ introduced in Eq. (1). For the flat surface in the experiments, the values of $\mu_f$ are extracted for the different liquids by adjusting the boundary condition $\mu_f$ in the axisymmetric CHNS simulations in order to match the evolution of the spreading radius and droplet profile of the corresponding experiments. We denote the obtained values as $\mu_{f\_flat}$, and consider them to be an intrinsic property of the combination of liquid and substrate surface chemistry, but independent of substrate geometry. The same matching is performed for the structured surfaces in the experiments, and the $\mu_f$ obtained here additionally incorporates the macroscopic effect of the substrate geometry, denoted as $\mu_{f\_eff}$ henceforth.

Fig. 3(a) shows examples of this procedure for the 1.0 cP liquid. The dashed solid line shows the result of the simulation for $\mu_f = 0$. The experiment on the flat substrate (red circles) is seen to be about 30% slower than this, but matches nicely with the simulation for $\mu_{f\_flat} = 0.090$ Pas. The experiments for the three structured surfaces in the graph are seen to be still slower, with the densest ($a=1.5$ μm, $b=0.5$ μm) showing the slowest spreading. For each structured surface and liquid an effective line friction coefficient $\mu_{f\_eff}$ is determined as described above. The droplet profile during spreading was also compared between simulation and experiment to assure the relevance of the analysis [Fig. 3(b)].

In order to isolate the importance of the surface geometry, we plot in Fig. 3(c) $\mu_{f\_eff}$ normalized by the corresponding $\mu_{f\_flat}$ as a function of the roughness parameter $S$. If we consider $\mu_{f\_flat}$ to represent the intrinsic surface chemistry of the liquid and solid system, then $\mu_{f\_eff}/\mu_{f\_flat}$ represents the geometric effects. The result in Fig. 3(c) quantifies that the net energy dissipation due to the microstructure increases with the roughness. We also note that the effect of this dissipation is significant and that it determines the spreading speed for our experiments.

One candidate for this mechanism of dissipation could be the work of internal viscous stresses in the complicated viscous flow around the pillars, and in the grooves between them. However, our analysis shows that for this to become important, a measured $\mu_{f\_eff}$ needs to be of the same order of magnitude as the liquid viscosity $\mu$. This is not the case and instead $\mu_{f\_eff}$ is between forty and several hundred times larger than $\mu$ for our experiments. We must therefore conclude that the added viscous dissipation for the flow through the microstructures is negligible compared to the line friction dissipation.

The other candidate for dissipation is obviously the line friction. As seen in Fig. 3(c), the behavior for the three viscous liquids is similar, with a 5-fold increase of $\mu_{f\_eff}/\mu_{f\_flat}$ as $S$ increases from 1 to 3. The values for the lowest viscosity are somewhat smaller, saturating at around 3 for values of $S$ above 2. This dependency on $S$ can be rationalized if we assume that the total dissipation rate is proportional to the total area that the contact line on average sweeps over per unit time, and that it on the microscopic level is given by the intrinsic $\mu_\text{f\_flat}$. This argument leads to a relation between $\mu_\text{f\_eff}$ and $\mu_\text{f\_flat}$ such that $\mu_\text{f\_eff}(a+b)^2=\mu_\text{f\_flat}\{(a+b)^2+4ah\}$, or $\mu_\text{f\_eff}/\mu_\text{f\_flat}=S$. As shown with the dashed line in Fig 3(c), this gives a dependency of $\mu_\text{f\_eff}/\mu_\text{f\_flat}$ on $S$ that is of the same magnitude as the experiments, while it does underestimate the line friction somewhat.

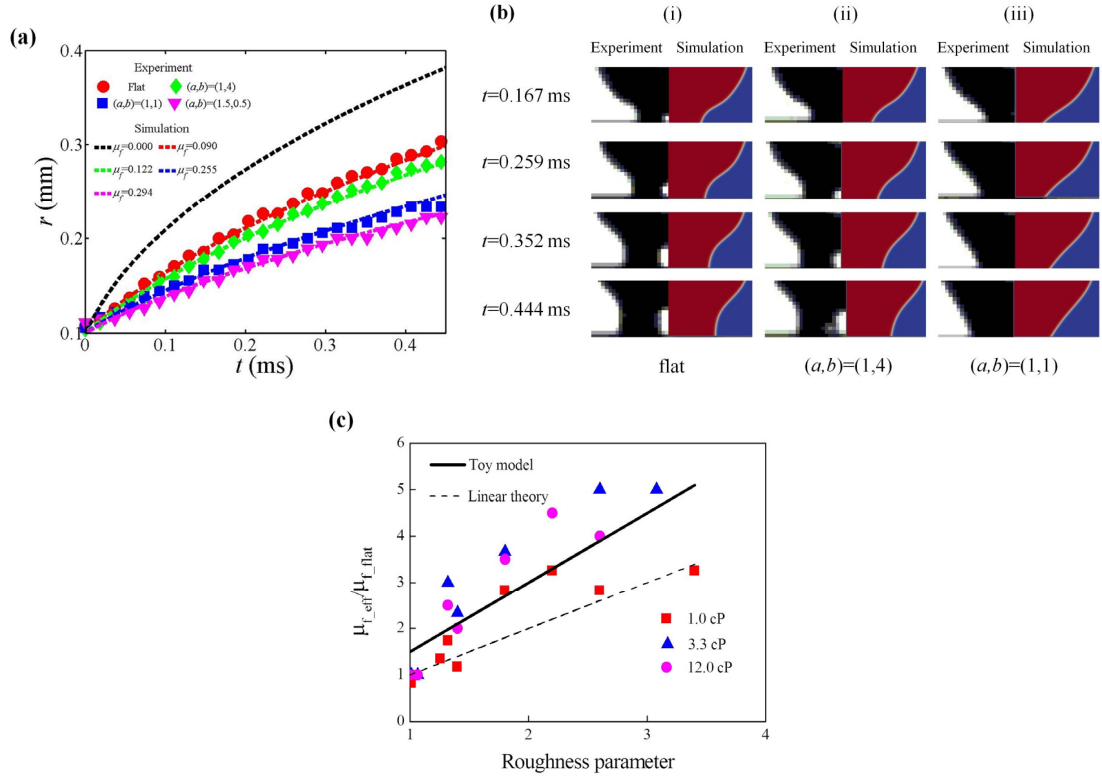

Figure 3 (a) Spreading radius as measured by experiment (markers) and simulation (dashed lines) tuned to match experiment by adjustment of $\mu_\text{f}$. (b) Confirmation of matching between experiment and simulation droplet profiles for various viscosities and patterns. (c) Dependence of normalized contact line friction $\mu_\text{f\_eff}/\mu_\text{f\_flat}$ on the roughness parameter $S =1+4ah/(a+b)^2$.

To study in detail how the flow around, and the advancement of the contact line over the pillars, give rise to the dissipation, we have performed CHNS simulations of a periodic cylindrical droplet spreading over a periodic 3D microscructure, with physical properties matched to those in the experiments. Figure 4(a) shows a snapshot of the droplet profile near the surface at different times for a representative case of $(a, b)=(3, 1)$. The yellow plane is inserted to show the profile of the droplet in case of the flat surface at $t=3.6$ μs, and the comparison clearly shows that the wetting is hindered by the microstructures. The simulation reveals the microscopic picture of wetting around the microstructure, where the contact line is seamlessly in contact with all the circumference of the microstucture and the length of the miscoscopic contact line is roughly proportional to the surface area.

Figure 4(a) also shows that the contact line is briefly pinned on corners in the structure, resulting in unsteady propagation of the contact line. The unsteady characteristics of the contact line can be better illustrated by plotting the position of the contact line along the center of the microstructures ["measurement line" in Fig. 4(a)]. As shown in Fig. 4(b), the propagation speed of the contact line indeed fluctuates with a period corresponding to that of the passage of the microstructure [denoted by the dashed lines on the sides of Fig. 4(b)]. The time histories of the contact line in the figure also confirms that the spreading is slower for microstructures with larger roughness parameter $S$.

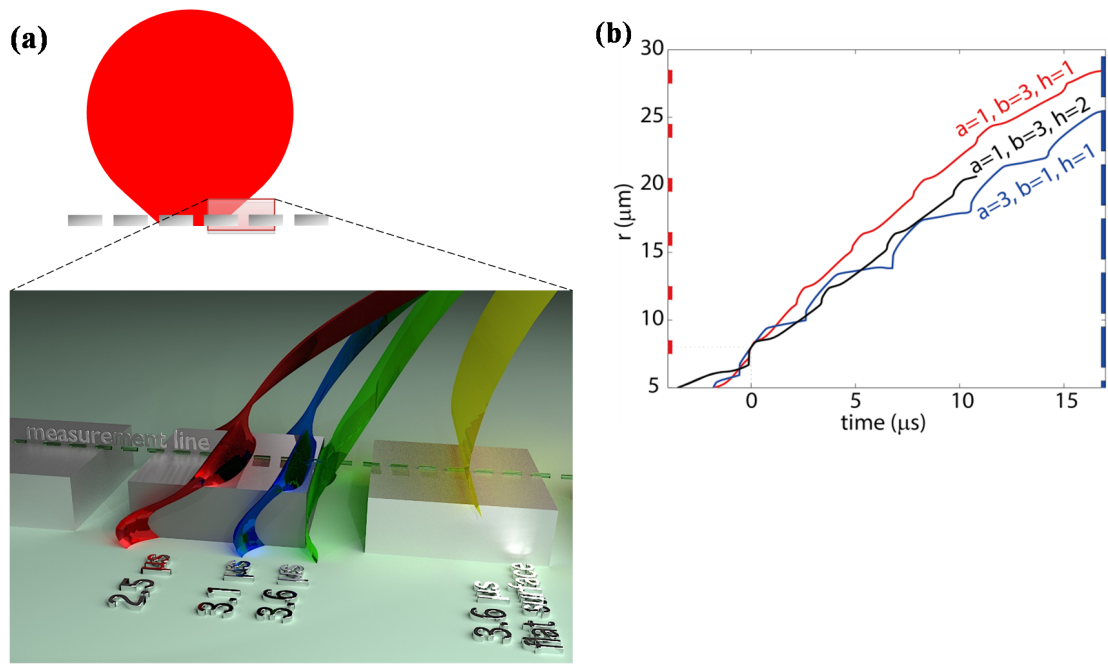

Figure 4 (a) Snapshot of the CHNS simulation for 2-D (cylindrical) droplet spreading on 3-D microstructures, moving from left to right. The red, blue, and green bands are the spreading front of the dropt on microstructure at times 2.5, 3.1, and 3.6 μs respectively from the beginning of wetting. For reference, the wetting front for the case of a flat surface is also shown (yellow band). (b) Simulated droplet radius as a function of time, highlighting the oscillation in velocity corresponding to the positions of the pillars, shown on the left side of the graph for ($a$, $b$, $h$)=(1, 3, 1) and on the right side for ($a$, $b$, $h$)=(3, 1, 1).

The strongly fluctuating character of the microscopic contact line motion suggests a mechanism for enhancement of the dissipation; with a velocity dependent friction it is most economical to keep a constant speed, and any fluctuations will increase dissipation. To estimate the importance of this, we note that the overall dissipation rate is given as $\mu_{f\_eff}U^2$, where $U$ is a spreading speed. On the microscopic level however we could consider the overall dissipation rate as a time average over a fluctuating spreading speed, $\mu_{f\_flat}u(t)^2$. Assuming that $u(t)=U\{1+\sin(\omega t)\}$, [a contact line speed fluctuating between zero and twice the mean speed, as seems representative of the profiles in Fig. 4(b)] we find that the time average is $(3/2)\mu_{f\_flat}U^2$. The fluctuations would thus cause an increase in the overall $\mu_{f\_eff}$ by a factor of 3/2, which gives the relation $\mu_{f\_eff}/\mu_{f\_flat}=(3/2)S$. This is shown in Fig. 3(c) as the solid straight line, and is seen to better capture the overall tendency in the experiment. Thus the general effective line friction coefficient can be denoted as

$$\mu^*_{f\_eff} = \frac{3}{2}\mu_{f\_flat}S \ . \tag{2}$$

The above knowledge regarding the dissipation mechanism helps us identify the velocity scale of spreading on structured surface. Figure 5(a) shows the Weber number We= $\rho R U^2/\gamma$ versus $\mathrm{Re}_\mu = \rho U R / \mu^*_{f\_eff}$. The Weber number is chosen as the appropriate nondimensional measure of velocity. $\mathrm{Re}_\mu$ has the form of a Reynolds number, but it should be noted that the viscosity here is the effective line friction [Eq. (2)]. As seen in Fig. 5(a), $\mathrm{Re}_\mu$ spans two decades, and over this range We is almost linear, with values ranging from 0.05 to around 3. A line can be fitted to the data points, which gives a relation We=$2.13\mathrm{Re}_\mu^{0.91}$. The almost linear dependence suggests that a capillary number defined as $\mathrm{Ca}_\mu$=We/$\mathrm{Re}_\mu$= $\mu^*_{f\_eff}U/\gamma$ should be of interest. As shown in Fig. 5(b), $\mathrm{Ca}_\mu$ indeed stays between 1.8 and 3.5 for the entire range of $\mathrm{Re}_\mu$. From the fit of We, we obtain that $\mathrm{Ca}_\mu$= $2.13\mathrm{Re}_\mu^{-0.095}$, i.e. a very slight decrease with increasing $\mathrm{Re}_\mu$. The fact that this capillary number stays of order unity for all our experiments strongly suggests that the spreading here is determined by the balance of the effective line friction and the driving force in terms of surface energy.

The main terms in the overall energy budget of the droplet evolution would be the acceleration of the liquid, the line friction as discussed above, and the driving force, i.e. the surface energies of the interfaces. Estimating the magnitude of these, the characteristic rate of change of inertia is $\rho U^3 R^2$, the dissipation at the contact line $\mu^*_{f\_eff}U^2R$, and the work of the surface tension $\gamma UR$. Normalizing these three terms by the driving force, we have the relative nondimensional estimates We, $\mathrm{Ca}_\mu$ and 1, respectively. Considering the dominant balance of these terms, the driving force has to balance either the inertia, or the line friction term. In case the contact line dissipation dominates, we obtain $Ca_\mu \sim 1$ and $U \sim \gamma/\mu^*_{f\_eff}$, showing a strong dependency on surface properties as discussed above. In the opposite case when inertia is dominating we obtain $We \sim 1$ and $U \sim (\gamma/\rho R)^{1/2}$, which in fact recovers the inertial time scale $\tau \sim (\rho R^3/\gamma)^{1/2}$. We also note that the velocity scale here does not include any property of the interface other than the surface energies, so in this parameter range the spreading should be insensitive to the features of the substrate.

In order to determine a priori which of the two situations we are to expect, we note that the condition for the crossover from the one parameter range to the other is obtained by having all three terms in the energy budget equal, $We \sim Ca_\mu \sim 1$. From this the velocity scale can be eliminated and we arrive at a condition for an Ohnesorge number based on the line friction coefficient: $Oh_\mu = (Ca_\mu^2/We)^{1/2} = (\mu^{*2}_{f\_eff}/\rho R\gamma)^{1/2}$. When $1/Oh_\mu \ll 1$, we expect line friction to dominate, i.e. $Ca_\mu \sim 1$, and hence, $U \sim \gamma/\mu^*_{f\_eff}$ and $We \sim (1/Oh)^{1/2} \ll 1$. In the opposite case, $1/Oh \gg 1$, we would have $We \sim 1$, $U \sim (\gamma/\rho R)^{1/2}$, and $Ca_\mu \sim Oh_\mu \ll 1$. This Ohnesorge number depends on the size of the droplet, the properties of the liquid and the surface and gives a criterion for determining the flow regime. We note that, for a small enough droplet, line friction will always dominate, and that conversely, a large enough droplet will always be inertial.

Figure 5(c) shows $We$ and $Ca_\mu$ as functions of $1/Oh_\mu$ for our data. We first notice that our $1/Oh_\mu$ are in a range from 0.01 to 1, i.e. it indeed covers the range where the line friction should dominate. The Weber number is also proportional to $Oh_\mu$ as expected in this range. Our data does not extend to $Oh \gg 1$, but we do expect that this range will recover the parameter range where the spreading is inertial and rather independent of the surface properties, as reported by other researchers[17-20]. Comparison with some $Oh > 1$ cases is shown in Fig. 5(c) by plotting the data extracted from previous experiments on smooth substrates with equilibrium contact angle similar to the current one[18,24,34]. The data agree well with the line fitted to our experiments for both $We$ and $Ca_\mu$, demonstrating the consistency with the above analysis.

In summary, we have found that the surface structure can greatly hinder the dynamic spreading of a partially-wetting droplet during the initial stages far from equilibrium. The spreading speed under typical conditions of surface roughness and liquid viscosity has been shown to be determined by dissipation associated with contact line friction. This line-friction is shown to dominate while the $\mu_{f\_flat} \gg \mu$, otherwise viscous dissipation dominates. The substrate structure contributes to this resistance chiefly by increasing the total area that is wetted as the contact line passes, while dissipation is further enhanced at the contact line by local unsteady motion associated with the rough surface. This should be present for any surface structure that shows hysteresis in contact angles. We derive an explicit expression for the effective line friction coefficient in terms of the intrinsic line friction coefficient and the geometrical parameters of the structure. The remaining contribution to the energy budget is the build-up of kinetic energy of the liquid in the droplet as it is accelerated by the capillary forces. We find that this dominates if an Ohnesorge number based on the line friction coefficient is less than unity, and in that case the spreading should be insensitive to the surface properties, while conversely surface properties will dominate for large Ohnesorge number. The results here reveal not only the underlying mechanism of how the geometry of the roughness influences the dynamic wetting, but also the possibility of selectively controlling the nonequilibrium wetting behavior by engineering the surface structure.

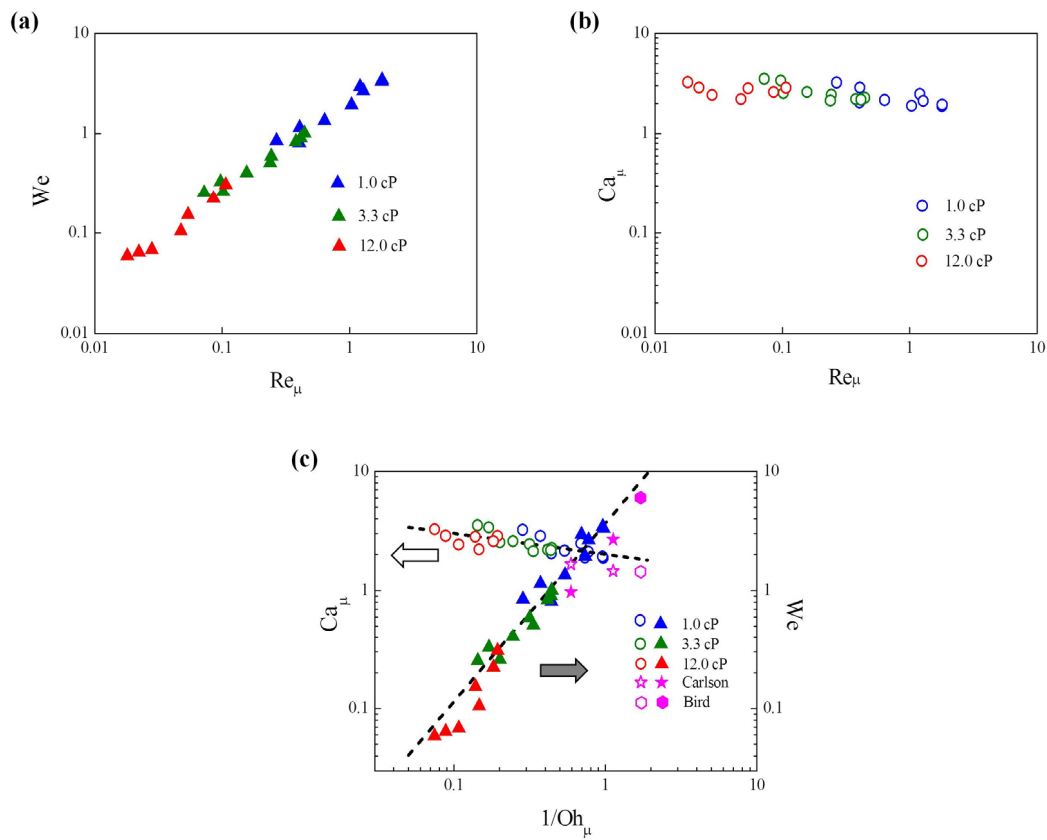

Figure 5 (a) Weber number We versus line friction Reynolds number $Re_\mu$. (b) Line friction capillary number $Ca_\mu$ versus line friction Reynolds number $Re_\mu$. (c) We and $Ca_\mu$ versus line friction Ohnesorge number $1/Oh_\mu$.

## Method

**Electron beam lithography.** HF (5 wt%) was used to remove the native silicon dioxide layer on the surface to get pure silicon wafer and 400 nm-thick ZEP-520 was spin-coated at 4000 rpm for 60 seconds. After being soft baked at 180 for 15 °C minutes, the wafer covered with photoresist was exposed with an EB lithography system (F5112+VD01, ADVANTEST) and the desired patterns were directly drawn on the substrate. Next, the wafer was immersed into ZED-N50 solution for 70 seconds for development and then moved to ZMD-B for cleaning. The patterns clearly emerged on the surface after the EB drawing process.

**Non-Bosch etching process.** The non-Bosch process was developed to be a suitable method for yielding vertical sidewalls of the pillars[35]. The adopted parameters were: gas rates of $SF_6$:$O_2$=50:50 sccm, pressure 3 Pa, platen power 10 W, coil power 1200 W, etching rate 1.8 μm/min and an etching time of 50s. As a result, the height of the pillars were all about 1.6 μm.

**Sample cleaning.** Sample cleaning was performed in order to obtain the microstructures with homogeneous surface chemistry. To start with, reactive-ion etching (RIE) was deployed to remove the remaining photoresist on the surface ($O_2$: 30 sccm, pressure: 26 Pa, RF: 100 W, Model: RIE-10NR, SAMCO). The microstructured sample was immersed in HF solution (5 wt%), and then exposed in air in a clean booth for three days to ensure that the surface was saturated with a native oxide layer, resulting in a stable and uniform surface.

**Selection of working liquids.** Ethanol was blended into water to lower the surface tension of the working liquids and glycerol was added to vary the viscosity. The three types of working liquids had nearly the same surface tension and the dynamic viscosities were, 1.0 cP, 3.3 cP, and 12.0 cP.

**Set up of spreading experiment.** For each trial, the desired substrate was placed under the thin needle (33G, 90°). The distance between the needle tip and the substrate was accurately fixed to be around 1.0 mm. As soon as the droplet came into contact with the surface, the rapid spreading process was captured and recorded by a Photron FASTCAM SA2 camera at 54 000 frames per second under a sufficiently intensive light provided by a Photron HVC-SL lamp. Following the experiments, the histories of spreading radius were extracted from the original snapshots through image processing.

**Measurement of average spreading rate.** The average spreading rates were calculated based on the observed histories of the spreading radii. The spreading radius was observed to develop linearly with time over a significant portion of each measurement, and therefore a linear least-squares fit was suitable to obtain the spreading rate. The spreading radius range over which this linear fit was applied was fixed for each value of viscosity, and was chosen to be appropriate for the range of radii observed as the roughness was varied.

**CNHS simulations.** The numerical simulations were carried out using femLego[36], a symbolic tool for solving partial differential equations using the adaptive finite element method, and utilizing the Cahn-Hillard equation, treated as a coupled system of the chemical potential and composition. An adaptively refined and derefined mesh has been used to ensure mesh resolution along the vicinity of the interface and the surface. In the axisymmetric CHNS simulations on flat surfaces, the effect of microstructures was incorporated through the contact line friction factor $\mu_f$. In the CHNS simulations of a 2D cylindrical droplet wetting the 3D microstructures, the actual geometries of the surface microstructure were resolved by adopting a model system of cylindrical-slice droplet, with the slice width of $2(a+b)$, and domain height and width of 3 and 2 times that of the droplet radius, respectively.

**References**

1   Lee, H., Lee, B. P. & Messersmith, P. B. A reversible wet/dry adhesive inspired by mussels and geckos. *Nature* **448**, 338-U334 (2007).

2   Blossey, R. Self-cleaning surfaces - virtual realities. *Nat. Mater.* **2**, 301-306 (2003).

3   Duez, C., Ybert, C., Clanet, C. & Bocquet, L. Making a splash with water repellency. *Nat. Phys.* **3**, 180-183 (2007).

4   Sirringhaus, H. *et al.* High-resolution inkjet printing of all-polymer transistor circuits. *Science* **290**, 2123-2126 (2000).

5   van Osch, T. H. J., Perelaer, J., de Laat, A. W. M. & Schubert, U. S. Inkjet printing of narrow conductive tracks on untreated polymeric substrates. *Adv. Mater.* **20**, 343-345 (2008).

6   Ahn, B. Y. *et al.* Omnidirectional Printing of Flexible, Stretchable, and Spanning Silver Microelectrodes. *Science* **323**, 1590-1593 (2009).

7   Onses, M. S. *et al.* Hierarchical patterns of three-dimensional block-copolymer films formed by electrohydrodynamic jet printing and self-assembly. *Nat. Nanotechnol.* **8**, 667-675 (2013).

8   Kwon, H.-M., Bird, J. C. & Varanasi, K. K. Increasing Leidenfrost point using micro-nano hierarchical surface structures. *Appl. Phys. Lett.* **103**, 201601 (2013).

9   Bird, J. C., Dhiman, R., Kwon, H.-M. & Varanasi, K. K. Reducing the contact time of a bouncing drop. *Nature* **503**, 385-389 (2013).

10  Fan, H. Y. *et al.* Rapid prototyping of patterned functional nanostructures. *Nature* **405**, 56-60 (2000).

11  Park, J.-U. *et al.* High-resolution electrohydrodynamic jet printing. *Nat. Mater.* **6**, 782-789 (2007).

12  Stuart, M. A. C. *et al.* Emerging applications of stimuli-responsive polymer materials. *Nat. Mater.* **9**, 101-113 (2010).

13    Duprat, C., Protiere, S., Beebe, A. Y. & Stone, H. A. Wetting of flexible fibre arrays. *Nature* **482**, 510-513 (2012).

14    Galliker, P. *et al.* Direct printing of nanostructures by electrostatic autofocussing of ink nanodroplets. *Nat. Comm.* **3**, 890 (2012).

15    Paxson, A. T. & Varanasi, K. K. Self-similarity of contact line depinning from textured surfaces. *Nat. Comm.* **4**, 1492 (2013).

16    Tanner, L. H. Spreading of silicone oil drops on horizontal surfaces. *J. Phys. D: Appl. Phys.* **12**, 1473-1484 (1979).

17    Biance, A. L., Clanet, C. & Quere, D. First steps in the spreading of a liquid droplet. *Phys. Rev. E* **69**, 016301 (2004).

18    Bird, J. C., Mandre, S. & Stone, H. A. Short-time dynamics of partial wetting. *Phys. Rev. Lett.* **100**, 234501 (2008).

19    Winkels, K. G., Weijs, J. H., Eddi, A. & Snoeijer, J. H. Initial spreading of low-viscosity drops on partially wetting surfaces. *Phys. Rev. E* **85**, 055301(R) (2012).

20    Eddi, A., Winkels, K. G. & Snoeijer, J. H. Short time dynamics of viscous drop spreading. *Phys. Fluids* **25**, 013102 (2013).

21    de Gennes, P.-G., Brochard-Wyart, F. & Quéré, D. *Capillarity and Wetting Phenomena*. (Springer-Verlag, 2004).

22    Blake, T. D. The physics of moving wetting lines. *J. Colloid Interface Sci.* **299**, 1-13 (2006).

23    Bertrand, E., Blake, T. D. & De Coninck, J. Influence of solid-liquid interactions on dynamic wetting: a molecular dynamics study. *J. Phys.: Condens. Matter* **21**, 464124 (2009).

24    Carlson, A., Bellani, G. & Amberg, G. Universality in dynamic wetting dominated by contact-line friction. *Phys. Rev. E* **85**, 045302(R) (2012).

25    Carlson, A., Bellani, G. & Amberg, G. Contact line dissipation in short-time dynamic wetting. *Europhys. Lett.* **97**, 44004 (2012).

26    Cazabat, A. M. & Stuart, M. A. C. Dynamics of wetting - effects of surface-roughness. *J. Phys. Chem.* **90**, 5845-5849 (1986).

27    Yuan, Q. & Zhao, Y.-P. Multiscale dynamic wetting of a droplet on a lyophilic pillar-arrayed surface. *J. Fluid Mech.* **716**, 171-188 (2013).

28    Kusumaatmaja, H., Vrancken, R. J., Bastiaansen, C. W. M. & Yeomans, J. M. Anisotropic drop morphologies on corrugated surfaces. *Langmuir* **24**, 7299-7308 (2008).

29  Courbin, L., Bird, J. C., Reyssat, M. & Stone, H. A. Dynamics of wetting: from inertial spreading to viscous imbibition. *J. Phys.: Condens. Matter* **21**, 464127 (2009).

30  Kubiak, K. J., Wilson, M. C. T., Mathia, T. G. & Carras, S. Dynamics of Contact Line Motion During the Wetting of Rough Surfaces and Correlation With Topographical Surface Parameters. *Scanning* **33**, 370-377 (2011).

31  De Gennes, P. G. Weting-statics and dynamics. . *Rev. Mod. Phys*. **57**, 827-863 (1985).

32  Chu, K.-H., Xiao, R. & Wang, E. N. Uni-directional liquid spreading on asymmetric nanostructured surfaces. *Nat. Mater.* **9**, 413-417 (2010).

33  Li, X. Y., Mao, L. Q. & Ma, X. H. Dynamic Behavior of Water Droplet Impact on Microtextured Surfaces: The Effect of Geometrical Parameters on Anisotropic Wetting and the Maximum Spreading Diameter. *Langmuir* **29**, 1129-1138 (2013).

34  Carlson, A., Do-Quang, M. & Amberg, G. Dissipation in rapid wetting. *J. Fluid Mech.* **682**, 213-240 (2011).

35  Kirikae, D., Suzuki, Y. & Kasagi, N. A silicon microcavity selective emitter with smooth surfaces for thermophotovoltaic power generation. *J. Micromech. Microeng.* **20**, 104006 (2010).

36  Do-Quang, M., Villanueva, W., Singer-Loginova, I. & Amberg, G. Parallel adaptive computation of some time-dependent materials-related microstructural problems. *Bulletin of the Polish Academy of Sciences-Technical Sciences* **55**, 229-237 (2007).

**Acknowledgements**

This work was financially supported in part by, the Japan Society for the Promotion of Science (J. W., J. C., and J. S), Swedish Governmental Agency for Innovation Systems (M. D.-Q. and G. A.), and the Japan Science and Technology Agency through CREST (J. W, J. C, and J. S).